\documentstyle[12pt]{article}
\begin{document}
\begin{center}
CHAOS SYNCHRONIZATION BETWEEN TIME DELAY COUPLED JOSEPHSON JUNCTIONS GOVERNED BY A CENTRAL JUNCTION \\
E. M. Shahverdiev\footnote{Corresponding author's e-mail: shahverdiev@physics.ab.az}, L. H. Hashimova, P. A. Bayramov and R. A. Nuriev\\ 
Institute of Physics, 33 H. Javid Avenue, Baku, AZ1143, Azerbaijan\\
~\\
ABSTRACT\\
\end{center}
We study chaos synchronization between Josephson junctions governed by a central junction with a time delay and demonstrate with numerical simulations the possibility of high quality synchronization. The results are important for obtaining high power systems of Josephson junctions, which are promising for practical applications.
PACS number(s):05.45.Xt, 05.45 + b, 05.45.Gg, 74.50.+r,74.40.+k\\
~\\
\begin{center}
I.INTRODUCTION
\end{center}
\indent Josephson junctions are paradigmatic nonlinear dynamical systems, see e.g. references in [1-2]. The study of chaos and its control in such systems [3-6] could be of huge practical importance from the application point of view. Chaos control in such systems is important for Josephson junction devices such as voltage standards, detectors, SQUID (Superconducting Quantum Interference Device), etc. where chaotic instabilities are not desired. Chaotic Josephson junctions [7-8] can be used for short-distance secure communications in wi-fi systems, high-bandwidth telecommunications, ranging purposes, etc [9-10]. In a chaos-based secure communications a message is masked in the broadband chaotic output of the transmitter and synchronization between the transmitter and receiver systems is used to recover a transmitted message [3-6, 11-16].\\
\indent It is well-known that Josephson junctions are a source of radiation with frequencies up to the terahertz (THz) region [17-18]. Terahertz waves have a wide range of possible uses, including security scanning, remote sensing chemical signatures of explosives, non-invasive applications in medicine [9]. One of the principal challenges in the field is to develop compact, low-cost, efficient THz sources. Additionally, for many remote sensing and imaging applications, high optical powers are desirable, in part owing to the significant attenuation of THz radiation by water vapor in the atmosphere [19, 17-18]. Unfortunately the radiation from a single Josephson junction is very weak, of the order of pW. Synchronization of arrays of Josephson junctions is one of the ways to increase the radiation power from such sources [17, 19-20]. Upon achieving synchronization the radiation power will be proportional to the number of Josephson junctions squared. By synchronizing tens of thousands junctions radiation power as high as mW can be achieved. Recently it was established that certain highly anisotropic cuprate superconductors naturally realize a stack of strongly electromagnetically coupled intrinsic Josephson junctions [17].\\
\indent As shown in [21] the effect of the coupling delay time between junctions coupled in series on the synchronization quality is significant. As demonstrated by the numerical simulations in [21], with increasing coupling delays between the Josephson junctions correlation coefficients decay rapidly. The same tendency also holds for the increased number of junctions. However in [21] we were able to show that high quality synchronization is still a possibility among unidirectionally coupled in a series of 9 Josephson junctions on condition that the coupling delay between the junctions are reduced significantly (hundreds of times). It is clear that there is a physical limit to such a reduction. In light of this it is important to search for alternative ways of coupling between the junctions to synchronize very large number of the coupled systems. Results of numerical modelling presented in this paper show that coupling via a common medium-a central junction- could be one such viable alternatives.\\
\indent This paper is organized as follows. In section II we present a model of Josephson junctions coupled via a central junction with time delay. Section III deals with numerical simulations of the model under study and compares the result of simulations with the case of Josephson junctions coupled in series. Conclusions are presented in Section IV. Acknowledgements are presented in Section V.\\
\begin{center}
II.JOSEPHSON JUNCTIONS GOVERNED BY A CENTRAL JUNCTION
\end{center}
\indent Consider the following Josephson junctions $2,3,4$ governed with a time delay by a central junction $1$ (Figure 1):  
\begin{equation}
\frac{d^{2}\phi_{1}}{dt^{2}} + \beta \frac{d\phi_{1}}{dt} + \sin\phi_{1} =i_{dc} + i_{0}\cos(\Omega t+\theta )
\end{equation}

\begin{equation}
\frac{d^{2}\phi_{2}}{dt^{2}} + \beta \frac{d\phi_{2}}{dt} + \sin\phi_{2} =i_{dc} + i_{0}\cos(\Omega t +\theta )-\alpha(\frac{d\phi_{2}}{dt} -\frac{d\phi_{1}(t-\tau_{1})}{dt})
\end{equation}

\begin{equation}
\frac{d^{2}\phi_{3}}{dt^{2}} + \beta \frac{d\phi_{3}}{dt} + \sin\phi_{3} =i_{dc} + i_{0}\cos(\Omega t +\theta )-\alpha(\frac{d\phi_{3}}{dt} -\frac{d\phi_{1}(t-\tau_{2})}{dt})
\end{equation}

\begin{equation}
\frac{d^{2}\phi_{4}}{dt^{2}} + \beta \frac{d\phi_{4}}{dt} + \sin\phi_{4} =i_{dc} + i_{0}\cos(\Omega t +\theta )-\alpha(\frac{d\phi_{4}}{dt} -\frac{d\phi_{1}(t-\tau_{3})}{dt})
\end{equation}

Where $\phi_{1},\phi_{2},\phi_{3}$,and $\phi_{4}$ are the phase differences of the superconducting order parameter across the junctions $1,2,3$ and $4$, respectively;$\beta$ is called the damping parameter 
$(\beta R)^2=\hbar(2eI_{c}C)^{-1}$, where $I_{c}, R $ and $C $ are the junctions' critical current, the junction resistance, and capacitance, respectively;$\beta$ is related to McCumber parameter $\beta_{c}$:$\beta^{2}\beta_{c}=1$;
$\hbar$ is Planck's constant divided by $2\pi$;$e$ is the electronic charge;$i_{dc}$ is the driving the junctions direct current; It is worth noting that in general the total current passing through the junction will contain contributions  from supercurrent carried by the electron pairs, from a displacement current and an ordinary current. Representing the displacement current by a capacitor, and the ordinary current by a resistor one arrives at the equivalent circuit for Josephson junction shown in Figure 2. This junction circuit is a building block for the schematic circuits shown in Figure 1 and Figure 3.\\
$i_{0}\cos(\Omega t+\theta)$ is the driving ac (or rf) current with amplitudes 
$i_{0}$,frequencies $\Omega $ and phases $\theta $;$\tau$ is the coupling delay time between the junctions $1-2$, $1-3$ and $1-4$; coupling between the junctions $1-2$,$1-3$ and $1-4$ is due to the currents flowing through the coupling resistors $R_{s}$ between the junctions $1-2$,$1-3$ and $1-4$ ;$\alpha_{s}=R \beta R_{s}^{-1}$ is the coupling strength between the junctions $1$-$2$, $1$-$3$ and $1$ - $4$. We note that in Eqs.(1-4)direct current and ac current amplitudes are normalized with respect to the critical currents for the relative  Josephson junctions; ac current frequencies $\Omega $ are normalized with respect to the Josephson junction plasma frequency $\omega^{2}=2eI_{c}(\hbar C)^{-1},$ and dimensionless time is normalized to the inverse plasma frequency. \\
\indent We emphasize that Josephson junction $1$ is the central one (see Figure 1) and is connected to junctions $2, 3, 4$ unidirectionally through the coupling resistors $R_{s}$ with some delay. We also underline there is no direct connection between the junctions $2, 3, 4$ to be synchronized. This is a principal distinction from the connection topology considered in [21]. In 
[21] the Josephson junctions to be synchronized were connected unidirectionally in series (Figure 3). In Figures 1 and 3 unidirectionality of the coupling between the junctions is provided by a Ferrite Isolator-FI. As the information propagation speed is finite, any transmission (connection) line between the junctions will provide a delay time. \\
\indent It is noted that if the external drive current is purely dc then we have a second order autonomous system, Eq.(1) and therefore chaotic dynamic for Eq. (1) is ruled out. In order to make the dynamics of the Josephson junction (1) chaotic, one can add an ac current to the external driving, which makes Eq.(1) a second order non-autonomous system. Then treating  $\Omega t$ term as a new dynamical variable Eq.(1) (and consequently Eqs.(2-4)) can be rewritten as a system of third-order ordinary differential equations, which can behave chaotically.
\begin{equation}
\frac{d\phi_{1}}{dt}=\psi_{1}
\end{equation}
\begin{equation}
\frac{d\psi_{1}}{dt}= - \beta \psi_{1} - \sin\phi_{1} + i_{dc}+ i_{0}\cos \varphi_{1}
\end{equation}
\begin{equation}
\frac{d\varphi_{1}}{dt}=\Omega
\end{equation}
\begin{equation}
\frac{d\phi_{2}}{dt}=\psi_{2}
\end{equation}
\begin{equation}
\frac{d\psi_{2}}{dt}= - \beta \psi_{2} - \sin\phi_{2} + i_{dc} + i_{0}\cos \varphi_{2}
 -\alpha_{s}(\psi_{2} -\psi_{1}(t-\tau))
\end{equation}
\begin{equation}
\frac{d\varphi_{2}}{dt}=\Omega
\end{equation}
\begin{equation}
\frac{d\phi_{3}}{dt}=\psi_{3}
\end{equation}
\begin{equation}
\frac{d\psi_{3}}{dt}= - \beta \psi_{3} - \sin\phi_{3} + i_{dc} + i_{0}\cos \varphi_{3}
 -\alpha_{s}(\psi_{3} -\psi_{1}(t-\tau))
\end{equation}
\begin{equation}
\frac{d\varphi_{3}}{dt}=\Omega
\end{equation}

\begin{equation}
\frac{d\phi_{4}}{dt}=\psi_{4}
\end{equation}
\begin{equation}
\frac{d\psi_{4}}{dt}= - \beta \psi_{4} - \sin\phi_{4} + i_{dc} + i_{0}\cos \varphi_{3}
 -\alpha_{s}(\psi_{4} -\psi_{1}(t-\tau))
\end{equation}
\begin{equation}
\frac{d\varphi_{4}}{dt}=\Omega
\end{equation}
\indent It is worth noting that Eq.(1) is also used for modeling of a driven nonlinear pendulum, charge density waves with a torque and sinusoidal driving field [22-23].\\
\begin{center}
III.NUMERICAL SIMULATIONS
\end{center}
\indent First we demonstrate that three Josephson junctions governed with time delay by the central junction can be synchronized. We note that the total number of junctions is four: a central junction connected to the other three remaining junctions (Figure 1). This is done in order to compare the results of this paper with the result of numerical simulations of synchronization between three connected in series junctions (Figure 3). We study chaos synchronization between Eqs.(5-16) using the correlation coefficient $C$ [24] 
$$C(\Delta t)= \frac{<(x(t) - <x>)(y(t) - <y>)>}{\sqrt{<(x(t) - <x>)^2><(y(t) - <y>)^2>}}$$
where $x$ and $y$ are the outputs of the interacting systems; the brackets$<.>$
represent the time average; this coefficient indicates the quality of synchronization: $C=1$ means perfect synchronization.\\
We simulate Eqs.(5-16) for the following set of parameters:$\beta=0.25,i_{dc}=0.3,i_{0}=0.7,\Omega=0.6, \theta=0, \alpha_{s}=0.45, \tau=0.2.$ Parameter values used in the numerical simulations are taken from [19-20]. 
Figure 4 demonstrates the error dynamics between the junctions $2-4$ (governed by junction 1) $\psi_{4}-\psi_{2}.$ While $C_{24}$ indicates the correlation coefficient between $\psi_{2}$ and $\psi_{4}.$ 
These simulation results underline the high quality synchronization between three Josephson junctions $2-4$ with different initial conditions. Figure 4a shows chaotic dynamics of the central junction $\psi_{1} (t)$.\\
\indent For comparison we also reproduce results of numerical simulations from [21] demonstrating high quality synchronization between the end junctions coupled in series with a time delay. Figure 5 shows the error $\psi_{3}-\psi_{1}$ dynamics between Josephson junctions 1 and 3 for the following set of parameters:$\beta=0.25,i_{dc}=0.3,i_{0}=0.7,\Omega=0.6, \theta=0, \alpha_{s}=0.45, \tau=0.2.$$C_{13}$ indicates the correlation coefficient between $\psi_{1}$ and $\psi_{3}.$ \\
\indent As mentioned in Section 1 the effect of the coupling delay time between the junctions coupled in series on the synchronization quality is significant: with increasing coupling delays between the Josephson junctions correlation coefficients decay rapidly. The same tendency also holds for the increased number of junctions. However in [21] we were able to show that high quality synchronization is still a possibility among unidirectionally coupled in a series of 9 Josephson junctions on condition that the coupling delay between the junctions are reduced significantly from $\tau=0.2$ to $\tau=0.001$. Figure 6 pictures $\psi_{9}$ vs. 
$\psi_{1}$ for the set of parameters $\beta=0.25,i_{dc}=0.3,i_{0}=0.7,\Omega=0.6, \theta=0, \alpha_{s}=0.45, \tau=0.001.$$C_{19}$ is the correlation coefficient between variables $\psi_{9}$ and $\psi_{1}.$\\
\indent Clearly there is a physical limit to such a reduction. If we keep $\tau=0.2$ then the synchronization quality between the end junctions in an array of 9 junctions is relatively low. Figure 7 presents the phase portrait between the dynamical variables $\psi_{9}$ and $\psi_{1}$ for $\beta=0.25,i_{dc}=0.3,i_{0}=0.7,\Omega=0.6, \theta=0, \alpha_{s}=0.45, \tau=0.2.$$C_{19}$ is the correlation coefficient between variables $\psi_{9}$ and $\psi_{1}.$\\
\indent In the light of this, an alternative way of coupling between the junctions in order to synchronize a very large number of junctions investigated in this paper is of considerable importance. In Figure 8 we present the results of numerical simulations in an array of 30 Josephson junctions coupled via a common medium-a central junction (in total 31 junctions) and show high quality synchronization is possible between the junctions with random initial conditions. For Figure 8 the coupling delay times between the central and other junctions to be synchronized are equal to $\tau=0.2.$ The other parameter values are as in Figure 4. $C_{2-31}$ is the correlation coefficient between variables $\psi_{31}$ and $\psi_{2}.$ As pictured in Figure 9 high quality synchronization takes place even in the case of 100 junctions coupled with the central junction with time delays $\tau=0.2;$ other parameter values are as in Figure 4. The correlation coefficient $C_{2-101}$ between dynamic variables $\psi_{101}$ and $\psi_{2}$ is equal to 0.99.
\begin{center}
IV.CONCLUSIONS
\end{center}
\indent In summary, we have numerically studied chaos synchronization between time-delay coupled Josephson junctions governed by a central junction. We have demonstrated the possibility of high quality synchronization between such systems. We have shown that the topology of the connection between the junctions is a very important factor for the synchronization quality. 
For many application purposes, e.g. in security scanning, remote sensing of explosives, non-invasive medicine,  high power Josephson junctions are required. Synchronization of arrays of Josephson junctions is one way to increase the radiation power from such sources.\\
\begin{center}
V.ACKNOWLEDGEMENTS
\end{center}
The authors  gratefully acknowledge Prof. K.A. Shore  (Bangor University, UK)  for  comments  on  the manuscript.\\
\newpage
\begin{center}
FIGURE CAPTIONS
\end{center}
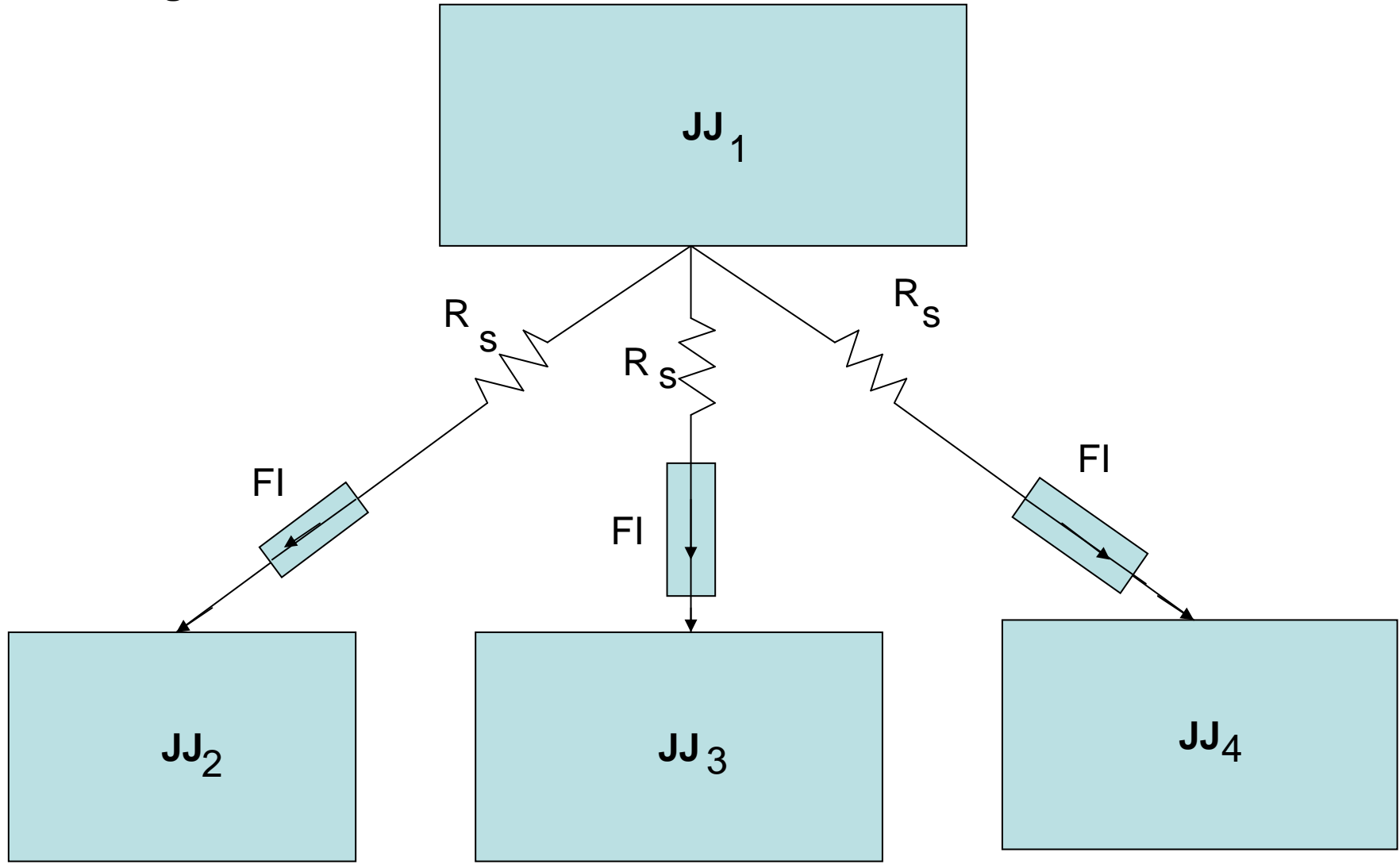
\noindent FIG.1.Schematic sketch of three Josephson junctions $2, 3, 4$ governed  with time delay by the central junction $1$.
Couplings 1-2, 1-3 and 1-4 occur with the same strength $\alpha_{s}$ via a resistor $R_{s}$. Unidirectionality of couplings is provided by a Ferric Isolator (FI).\\
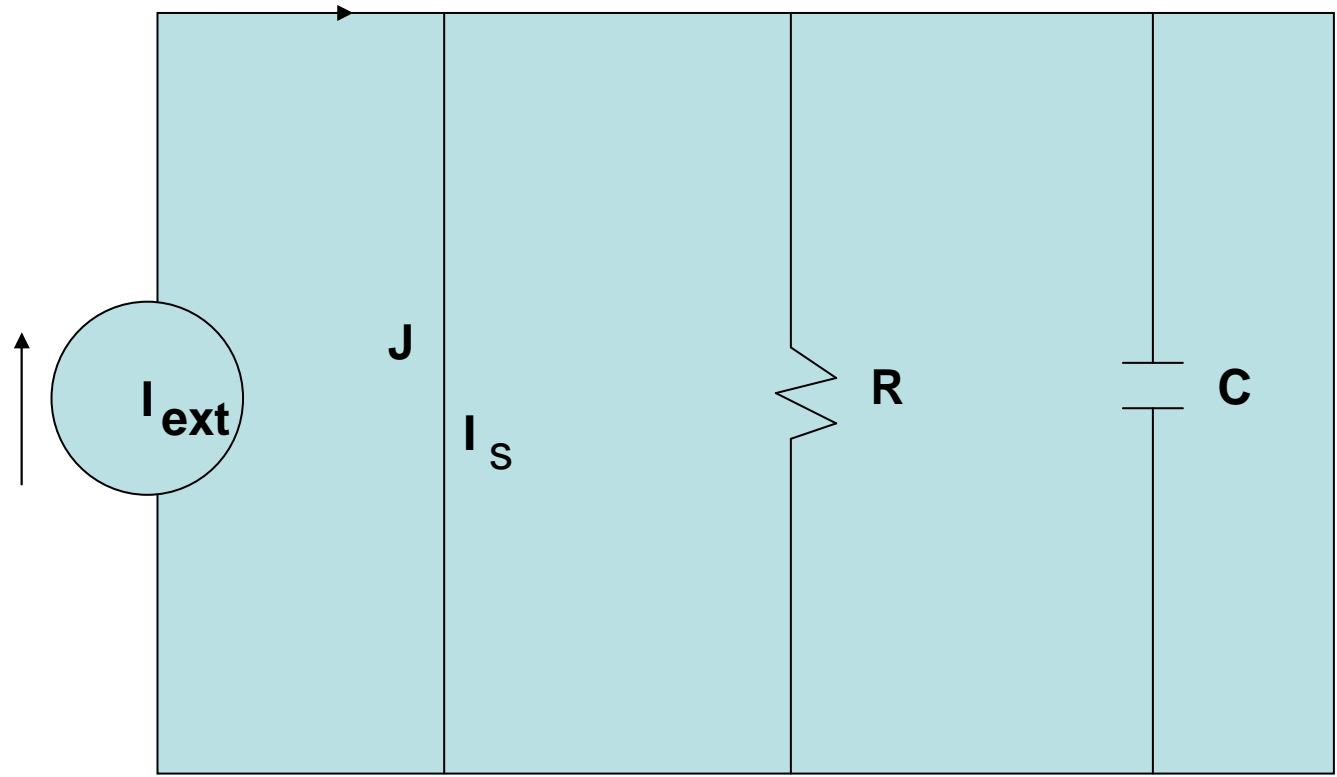
\noindent FIG.2.Schematic set-up of the RC-shunted Josephson Junction subject to the external driving $I_{ext}.$ The total current passing through the junction contains contributions from the supercurrent $I_{s}$ carried by electron pairs, the current through the capacitor $C$ and the current through the resistor $R$.\\
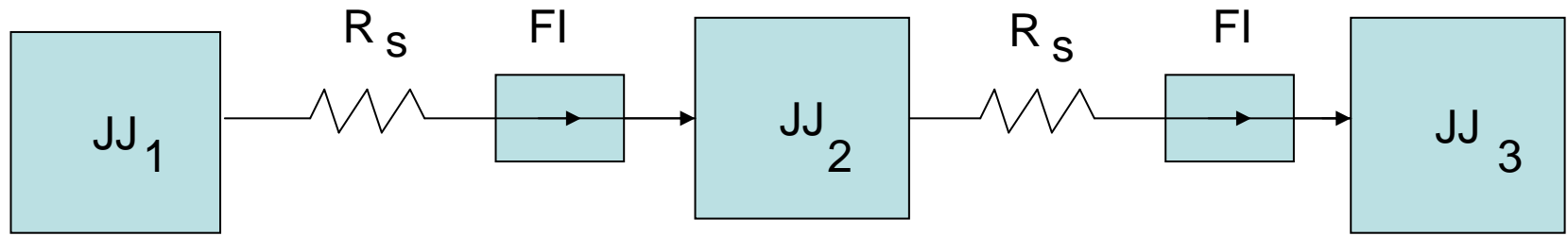
FIG.3.Schematic sketch of three Josephson junctions $1-2-3$ coupled in series. Unidirectional couplings 1-2 and 2-3 occur with the same strength $\alpha_{s}$ via a resistor $R_{s}$. Ferric Isolator (FI) provides unidirectionality of couplings.\\
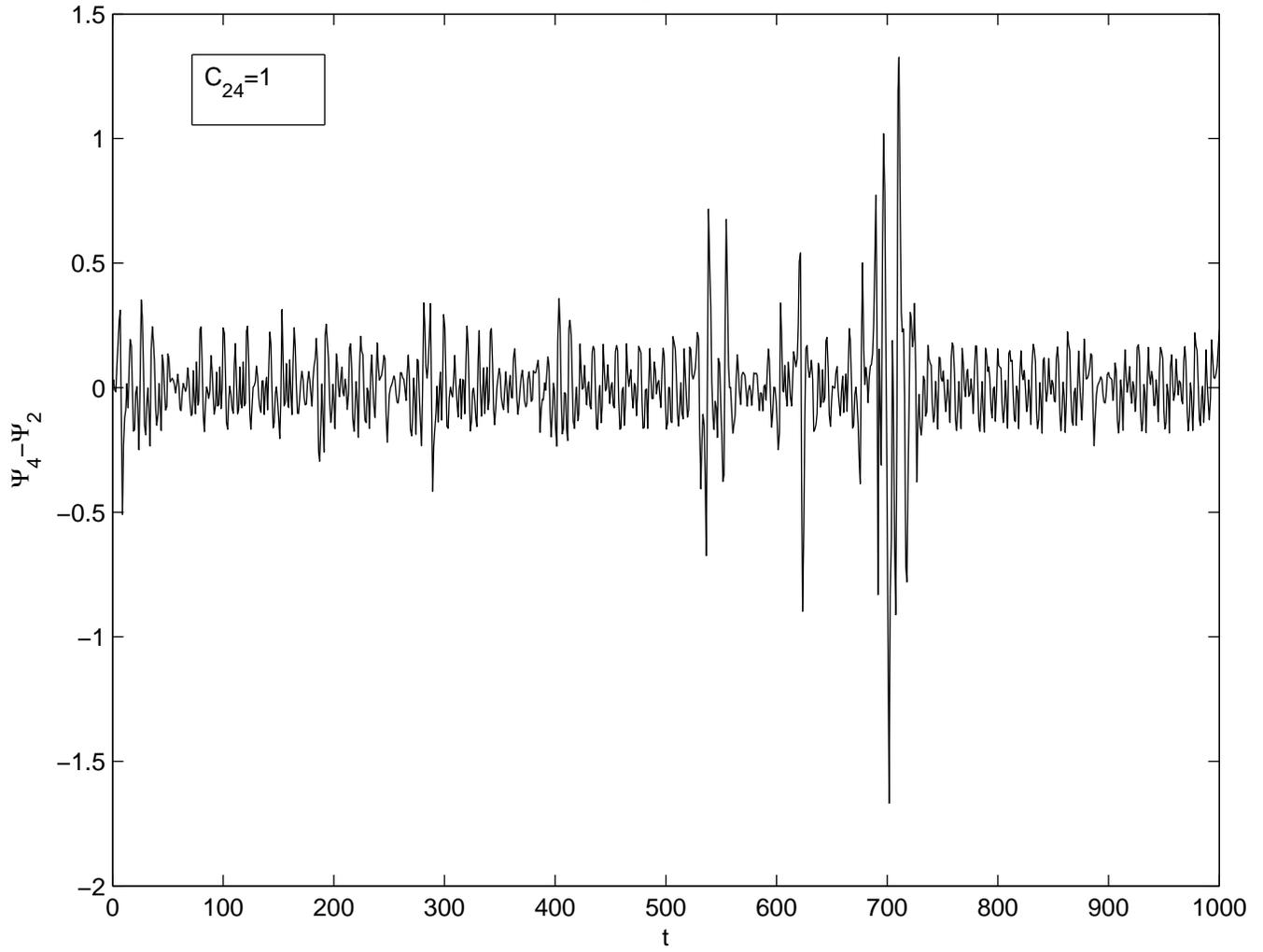
FIG.4.Numerical simulation of Eqs. (5-16): error dynamics of $\psi_{4}-\psi_{2}$:$C_{24}$ is the correlation coefficient between Josephson junctions 2 and 4. The parameters are:$\beta=0.25,i_{dc}=0.3,i_{0}=0.7,\Omega=0.6, \theta=0, \alpha_{s}=0.45, \tau=0.2.$Dimensionless units.\\
FIG.4a.Numerical simulation of Eqs.(5-7): Chaotic dynamics of the central junction $\psi_{1}(t)$.
The parameters are:$\beta=0.25,i_{dc}=0.3,i_{0}=0.7,\Omega=0.6, \theta=0, \alpha_{s}=0.45 $.Dimensionless units.\\
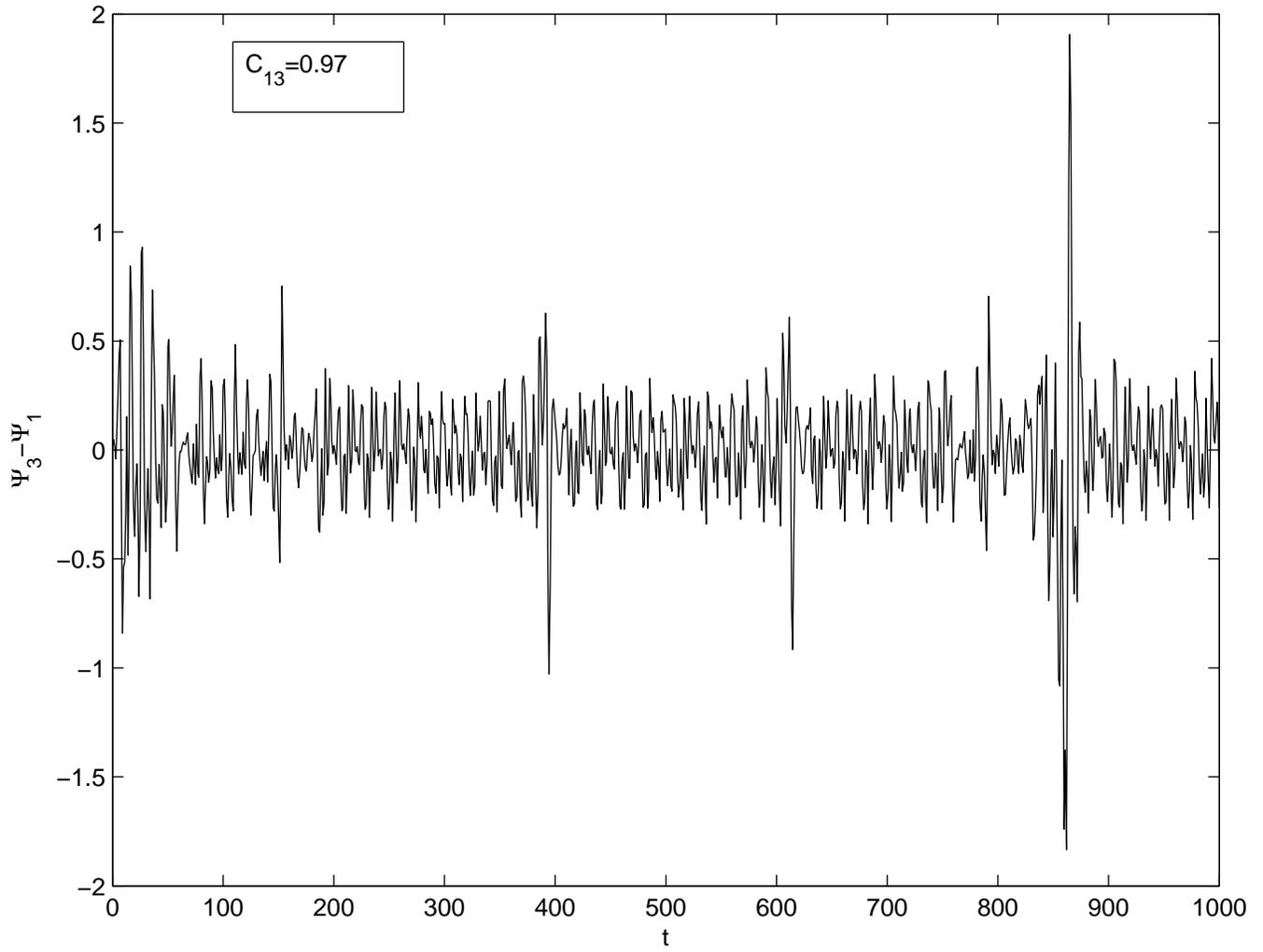
FIG.5.Numerical simulation of three Josephson junctions $1-2-3$ coupled in series: error dynamics of $\psi_{3}-\psi_{1}$:$C_{13}$ is the correlation coefficient between the end Josephson junctions 1 and 3. The parameters are:$\beta=0.25,i_{dc}=0.3,i_{0}=0.7,\Omega =0.6, \theta =0, \alpha_{s}=0.45, \tau =0.2.$Dimensionless units.\\
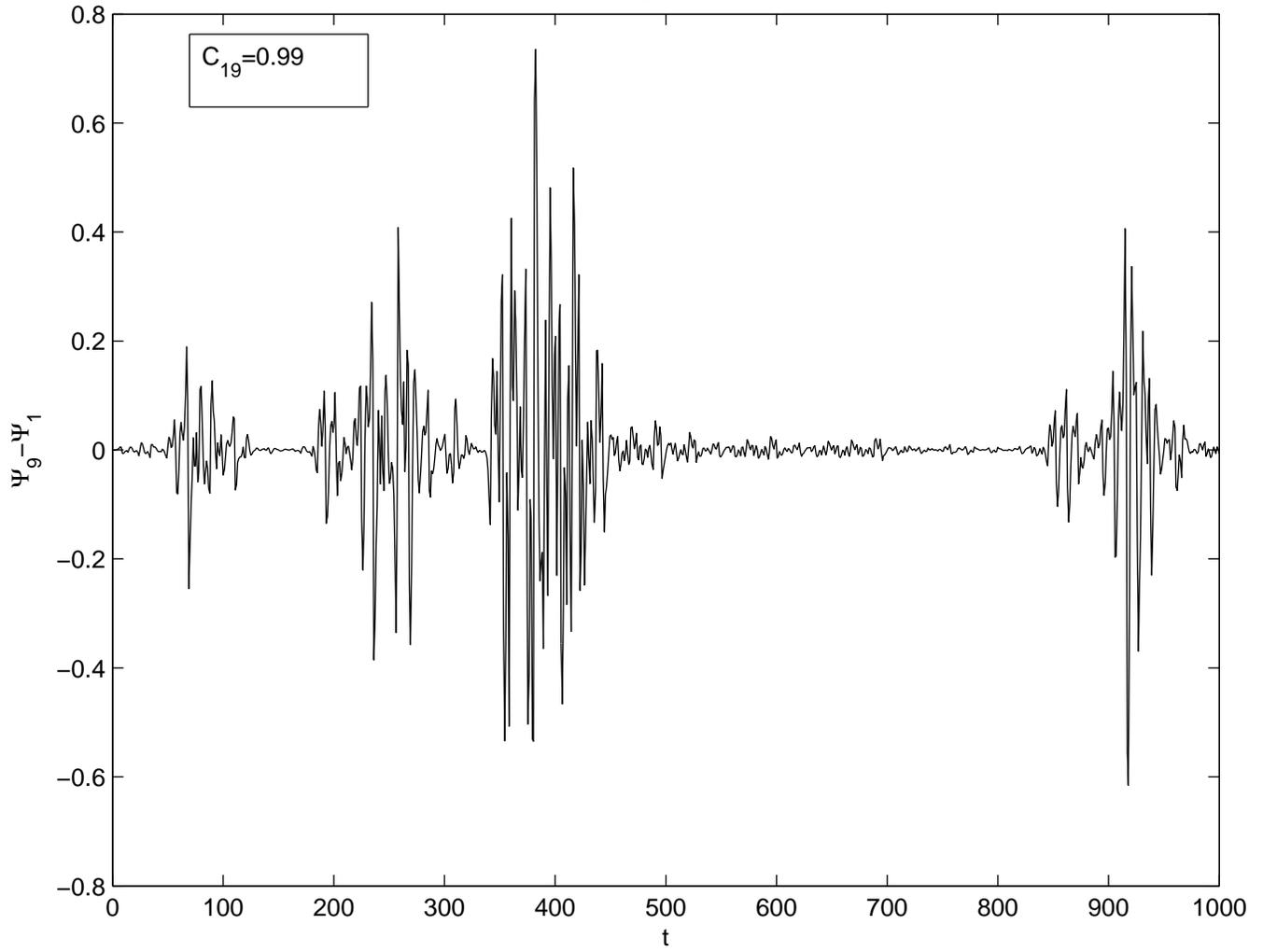
FIG.6.Numerical simulation of nine Josephson junctions coupled in series: $\psi_{9}$ vs.$\psi_{1}$:$C_{19}$ is the correlation coefficient between the end Josephson junctions 1 and 9. The parameters are: $\beta =0.25,i_{dc}=0.3,i_{0}=0.7,\Omega =0.6, \theta =0, \alpha_{s}=0.45, \tau =0.001.$Dimensionless units.\\
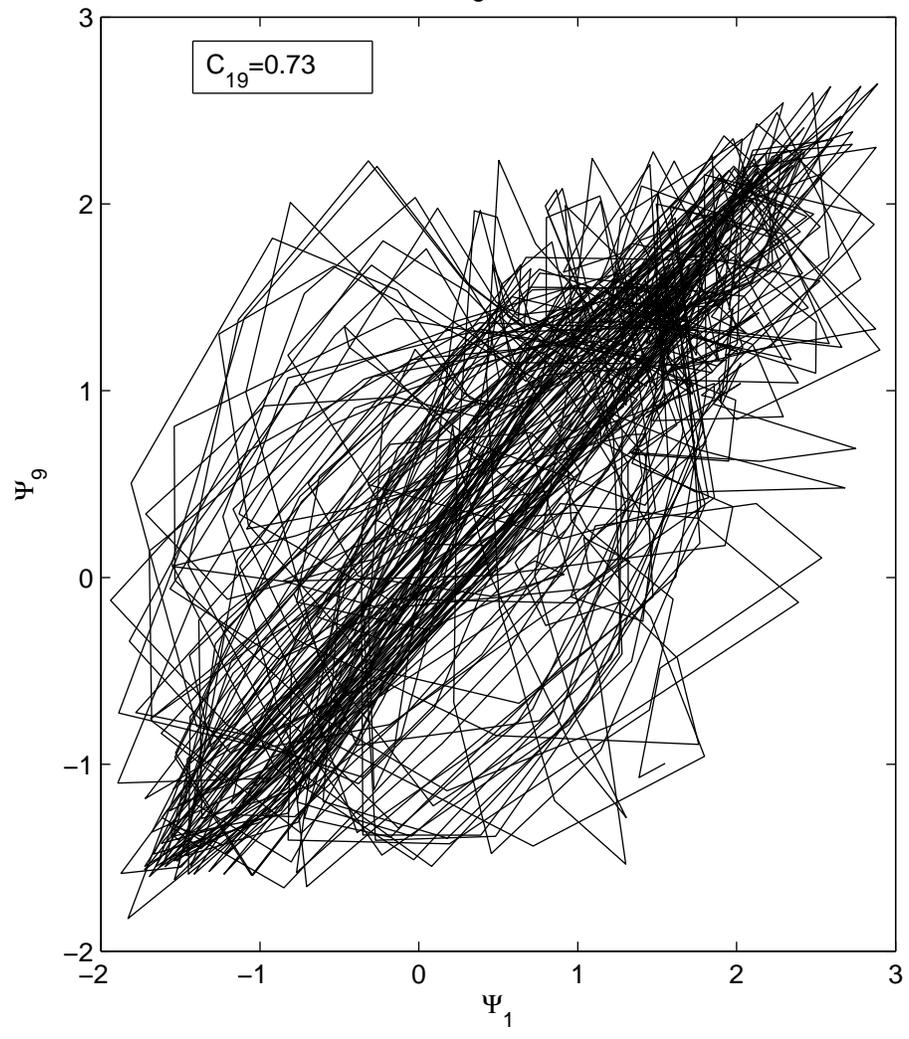
FIG.7.Numerical simulation of nine Josephson junctions coupled in series: Phase portrait between variables $\psi_{9}$ and$\psi_{1}$:$C_{19}$ is the correlation coefficient between the end Josephson junctions 1 and 9. The parameters are:$\beta =0.25,i_{dc}=0.3,i_{0}=0.7,\Omega=0.6, \theta =0, \alpha_{s}=0.45, \tau =0.2.$Dimensionless units.\\
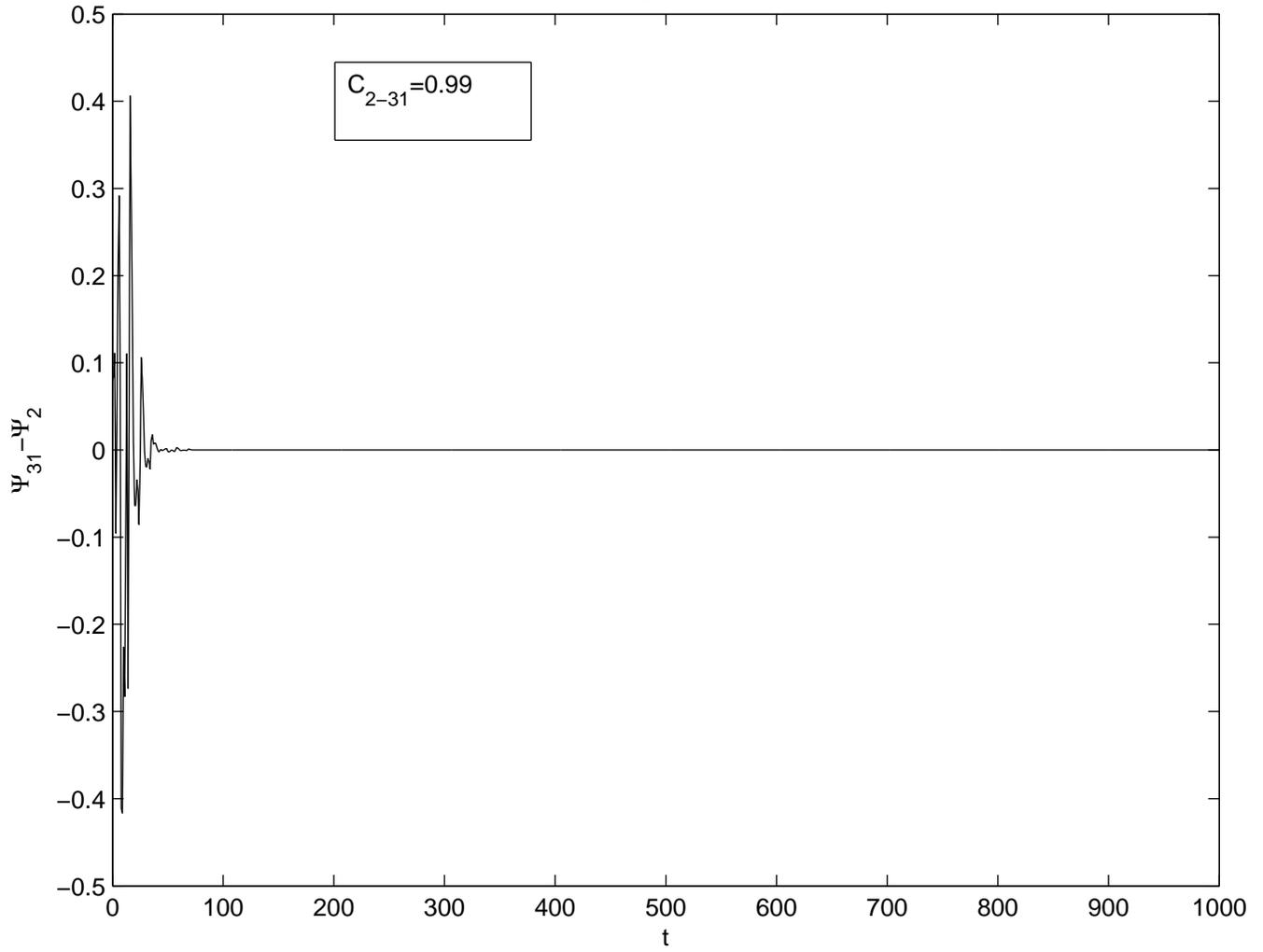
FIG.8.Numerical simulation of 30 Josephson junctions governed by a central junction: error dynamics of $\psi_{31}-\psi_{2}$:$C_{2-31}$ is the correlation coefficient between Josephson junctions 2 and 31. The parameters are: $\beta =0.25,i_{dc}=0.3,i_{0}=0.7,\Omega =0.6, \theta =0, \alpha =0.45, \tau =0.2$. Dimensionless units.\\
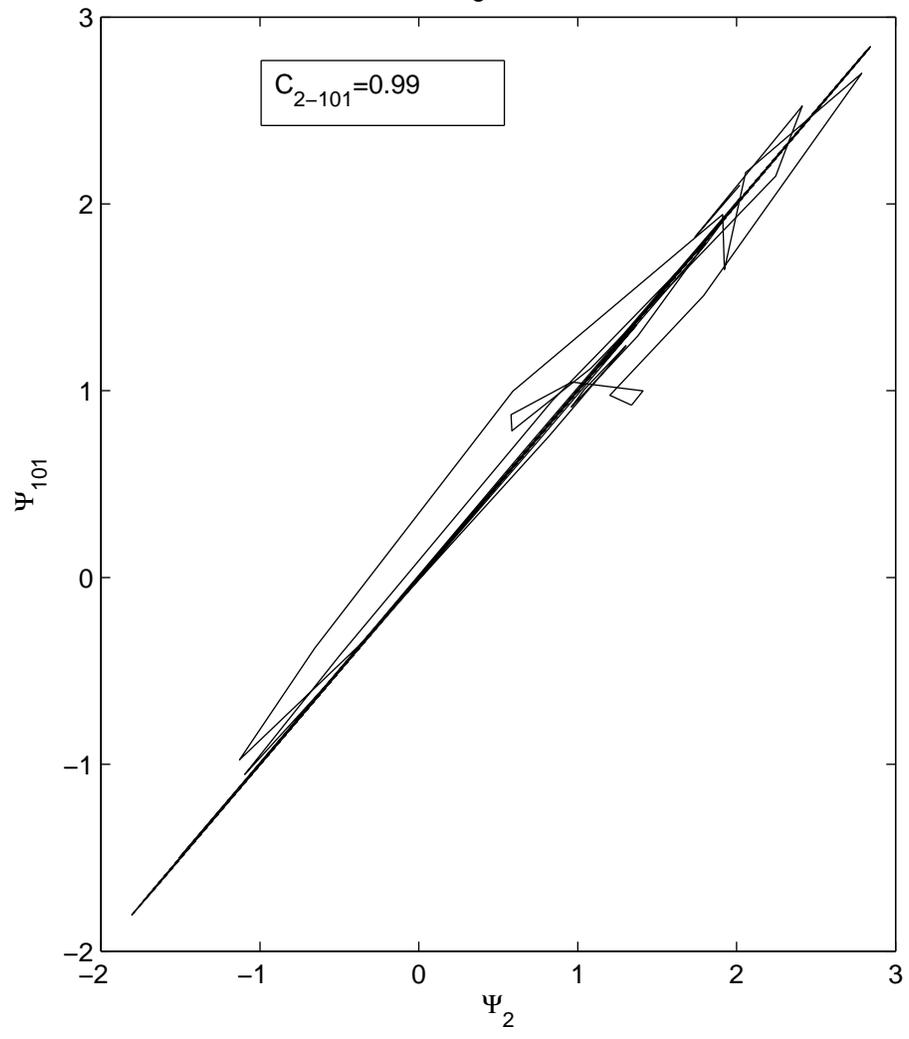
FIG.9.Numerical simulation of 100 Josephson junctions governed by a central junction: error dynamics of $\psi_{101}-\psi_{2}$:$C_{2-101}$ is the correlation coefficient between Josephson junctions 2 and 101. The parameters are: $\beta =0.25,i_{dc}=0.3,i_{0}=0.7,\Omega =0.6, \theta =0, \alpha_{s}=0.45, \tau =0.2.$Dimensionless units.\\

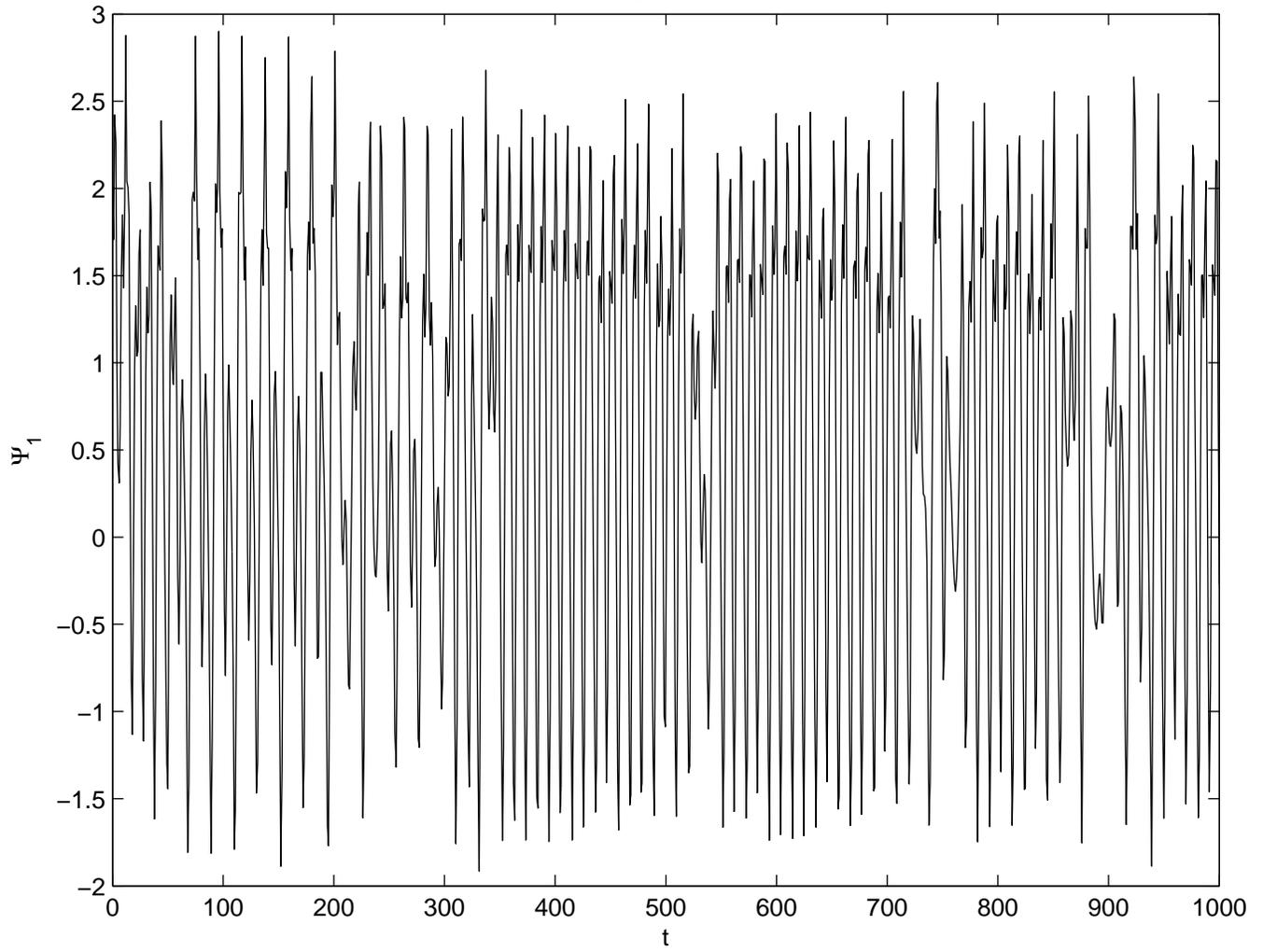
\end{document}